\documentclass[printer]{aa}
\usepackage{graphicx}
\usepackage{hyperref}
\hypersetup{draft}

\usepackage{natbib}
\bibpunct{(}{)}{;}{a}{}{,} 

\usepackage{txfonts}
\usepackage{longtable}
\usepackage{listings}
\usepackage{color}
\usepackage{textcomp}

\usepackage{subfig}

\begin{document}

\title{ A ring in a shell: the large-scale 6D structure of the Vela~OB2 complex \thanks{List of probable members and membership probabilities only available in electronic form
at the CDS via anonymous ftp to cdsarc.u-strasbg.fr (130.79.128.5)
or via http://cdsweb.u-strasbg.fr/cgi-bin/qcat?J/A+A/}}


\author{
T. Cantat-Gaudin\inst{\ref{IEECUB}}
\and
M. Mapelli\inst{\ref{IAT},\ref{OAPD}}
\and
L. Balaguer-N{\'u}{\~n}ez\inst{\ref{IEECUB}}
\and
C. Jordi\inst{\ref{IEECUB}}
\and
G. Sacco\inst{\ref{OAAR}}
\and
A. Vallenari\inst{\ref{OAPD}}
}

\institute{
Institut de Ci\`encies del Cosmos, Universitat de Barcelona (IEEC-UB), Mart\'i i Franqu\`es 1, E-08028 Barcelona, Spain\label{IEECUB}
\and
Institut f\"ur  Astro- und Teilchenphysik, Universit\"at Innsbruck, Technikerstrasse 25/8, A--6020, Innsbruck, Austria \label{IAT}
\and
INAF, Osservatorio Astronomico di Padova, vicolo Osservatorio 5, 35122 Padova, Italy\label{OAPD}
\and
INAF, Osservatorio Astrofisico di Arcetri, Largo E. Fermi 5, 50125, Florence\label{OAAR}
}

\date{Received date / Accepted date }

\abstract
{The Vela~OB2 association is a group of $\sim$10\,Myr stars exhibiting a complex spatial and kinematic substructure. The all-sky \textit{Gaia}~DR2 catalogue contains proper motions, parallaxes (a proxy for distance), and photometry that allow us to separate the various components of Vela~OB2.}
{We characterise the distribution of the Vela~OB2 stars on a large spatial scale, and study its internal kinematics and dynamic history.}
{We make use of \textit{Gaia}~DR2 astrometry and published \textit{Gaia}-ESO Survey data. We apply an unsupervised classification algorithm to determine groups of stars with common proper motions and parallaxes.}
{We find that the association is made up of a number of small groups, with a total current mass over 2330\,M$_{\odot}$. The three-dimensional distribution of these young stars trace the edge of the gas and dust structure known as the IRAS Vela Shell across $\sim$180\,pc and shows clear signs of expansion.}
{We propose a common history for Vela~OB2 and the IRAS Vela Shell. The event that caused the expansion of the shell happened before the Vela~OB2 stars formed, imprinted the expansion in the gas the stars formed from, and most likely triggered star formation.}


\maketitle{}

\section{Introduction}

The Vela~OB2 region is an association of young stars near the border between the Vela and Puppis constellations, first reported by \citet{Kapteyn14}. The association was known for decades as a sparse group of a dozen early-type stars \citep{Blaauw46,Brandt71,Straka73}. The modern definition of the Vela~OB2 association comes from the study of \citet{deZeeuw99}, who used Hipparcos parallaxes and proper motions to identify a coherent group of 93 members. Their positions on the sky overlaps with the IRAS Vela Shell \citep[][]{Sahu92phdt}, a cavity of relatively low gas and dust density, itself located in the southern region of the large Gum nebula \citep{Gum52}. 

The Vela~OB2 association is known to host the massive binary system $\gamma^2$~Vel, made up of a Wolf-Rayet (WR) and an O-type star;  recent interferometric distance estimates have located it at $336^{+8}_{-7}$\,pc \citep{North07}. \cite{deMarco99} estimate the total mass of the system to be 29.5$\pm$15.9\,M$_{\odot}$, while \citet{Eldridge09} propose initial masses of 35\,M$_{\odot}$ and 31.5\,M$_{\odot}$ for the WR and O components, respectively. \citet{Pozzo00}, making use of X-ray observations, identified a group of pre-main sequence (PMS) stars around $\gamma^2$~Vel, 350 to 400\,pc from us. This association is often referred to as the Gamma Velorum cluster, although \citet{Dias02} list it under the name Pozzo~1. \citet{Jeffries09} estimate that this cluster is 10\,Myr old. The \textit{Gaia}-ESO Survey study of \citet{Jeffries14} (hereafter J14) made use of spectroscopic observations to select the young stars on the basis of their Li abundances, and found the Gamma Velorum cluster to present a bimodal distribution in radial velocity. They suggested one group (population A) might be related to $\gamma^2$~Vel, while the other (population B) is not, and is in a supervirial state. The N-body simulations of \citet{Mapelli15} confirmed that given the small total mass of the Gamma Velorum cluster, the observed radial velocities could only be explained if both components were supervirial and expanding. In a study of the nearby 30\,Myr cluster NGC~2547, \citet{Sacco15} (hereafter S15) identified stars whose radial velocities matched that of population B of J14, 2$^{\circ}$ south of $\gamma^2$~Vel, hinting that this association of young stars might be very extended. \citet{Prisinzano16} estimate a total mass of the Gamma Velorum cluster of about 100\,M$_{\odot}$, which is a very low total mass for a group containing a system as massive as $\gamma^2$~Vel. As pointed out by J14, and according to the scaling relations of \citet{Weidner10}, the total initial mass of a cluster should be at least 800\,M$_{\odot}$ in order to form a 30\,M$_{\odot}$ star.

It has been suggested \citep[e.g.][]{Sushch11} that the IRAS Vela Shell (IVS) is a stellar-wind bubble created by the most massive of the Vela~OB2 stars. Observations of cometary globules \citep{Sridharan92,Sahu93,Rajagopal98} and gas \citep{Testori06} have shown that the IVS is currently in expansion at a velocity of $8-13$\,km\,s$^{-1}$. \citet{Testori06} point out that stellar winds from all the known O and B stars in the area are probably not sufficient to explain this expansion, and suggest it could have been initiated by a supernova, possibly the one from which the runaway B-type star HD~64760 (HIP~38518) originated \citep{Hoogerwerf01}. We note that \citet{Choudhury09} do not measure a significant expansion of the cometary globule distribution, but do observe that the tails of the 30 globules all point away from the centre of the IVS.

Based on \textit{Gaia}~DR1 data, \citet{Damiani17} have shown the presence of at least two dynamically distinct populations in the area. One of this groups includes the Gamma Velorum cluster, while the other includes NGC~2547. Still using \textit{Gaia}~DR1 data, \citet{Armstrong18} found that the young stars in the region are distributed in several groups and do not all cluster around $\gamma^2$~Vel.
Recently, \citet{Franciosini18} have used the astrometric parameters of the \textit{Gaia}~DR2 catalogue \citep{GDR2content} to confirm that populations A and B identified by J14 are located $\sim$38\,pc from each other along our line of sight, and observe a radial velocity gradient with distance which they interpret as a sign of expansion. The \textit{Gaia} DR2 catalogue was also used by \citet{Beccari18}, who find that the stars of the age of Gamma Velorum in a $10^{\circ}\times5^{\circ}$ field are not distributed in only two subgroups, but in at least four different subgroups.

In this study we use \textit{Gaia}~DR2 data to characterise the spatial distribution and kinematics of stars of the age of the Vela OB2 association.
This paper is organised as follows. Section~\ref{sec:data} presents the data and our source selection. Section~\ref{sec:components} contains a description of the observed spatial distribution. Section~\ref{sec:mass} discusses the mass content in the Vela OB2 complex. Section~\ref{sec:IVS} relates our findings to the context of the IVS. Sections~\ref{sec:discussion} discusses the results. Concluding remarks are presented in Sect.~\ref{sec:conclusion}.

\section{Data and target selection} \label{sec:data}
We obtained data from the \textit{Gaia}~DR2 archive\footnote{\url{https://gea.esac.esa.int/archive/}} on a larger area than previous studies ($\sim$30$^{\circ}\times30^{\circ}$) and in the parallax range $1.8 < \varpi < 3.4$\,mas, including the Gamma Velorum complex and nearby clusters. We follow the recommendations given in equations (1) and (2) of \citet{Arenou18} to keep sources with good astrometric solutions. The spatial distribution and proper motions of these stars are shown in Fig.~\ref{fig:fourpanels_regionmap}, along with the footprint of the areas covered by previous studies. We also indicate the positions and mean astrometric parameters of the nearby OCs, taken from \citet{CantatGaudin18gdr2}. In order to retain the stars related to the Gamma Velorum cluster, we first apply broad proper motion cuts, then perform a photometric selection.

\subsection{Proper motion pre-selection} \label{sec:propermotionselection}

\begin{figure*}[ht]
\begin{center} \resizebox{\hsize}{!}{\includegraphics[scale=0.5]{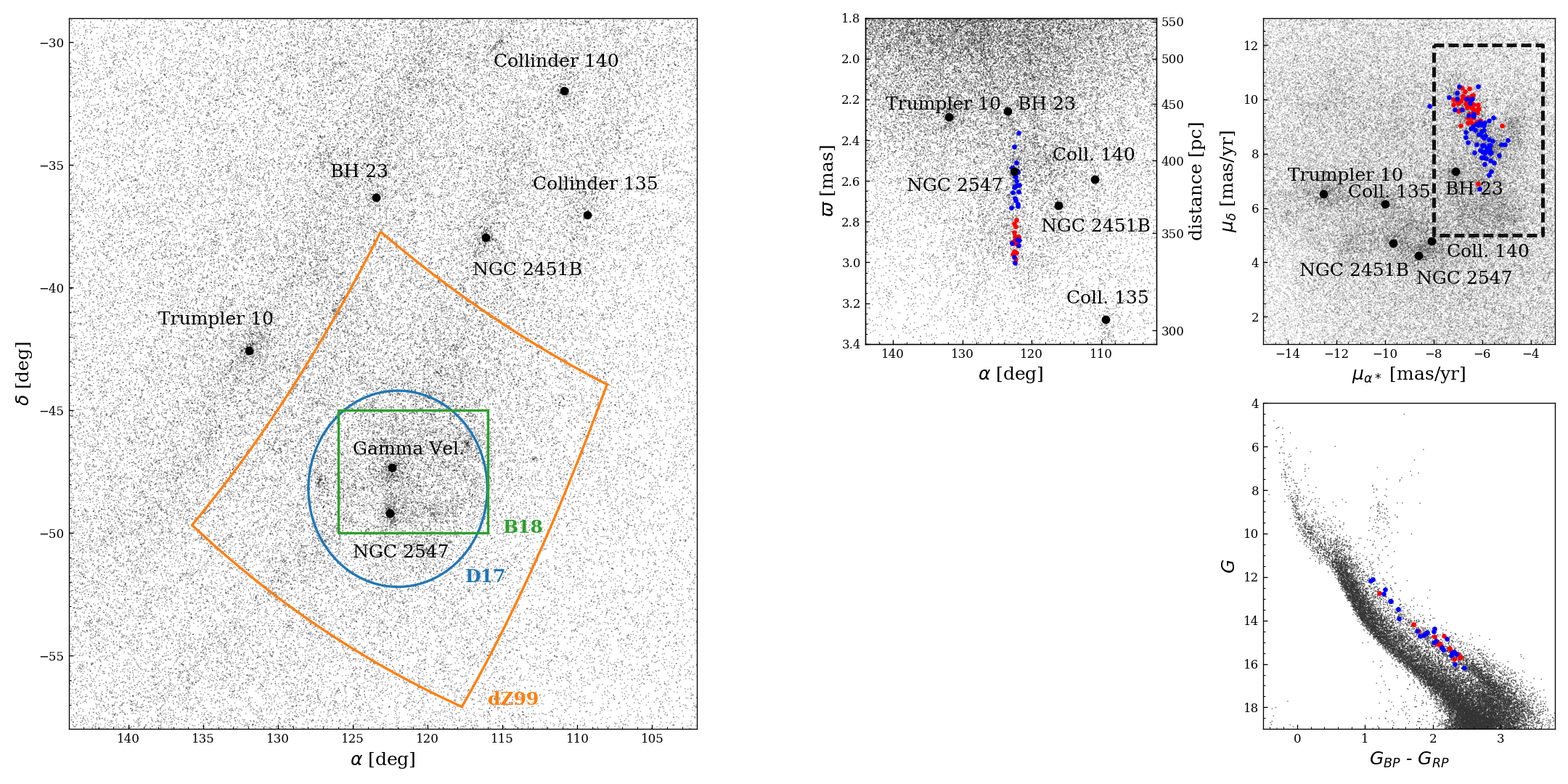}} \caption{\label{fig:fourpanels_regionmap} Left: Spatial distribution of \textit{Gaia} DR2 sources with parallaxes in the range $1.8 < \varpi < 3.4$\,mas. The black dots indicate known OCs in the area and distance range. We indicate the areas surveyed by dZ99 \citep{deZeeuw99}, D17 \citep{Damiani17}, and B18 \citep{Beccari18}.
Top middle: $\alpha$ vs $\varpi$.
Top right: Proper motion diagram for the same sources, indicating the mean proper motion of the known OCs \citep[taken from][]{CantatGaudin18gdr2}. The dashed box indicates our initial proper motion selection. Red and blue points respectively indicate populations A and B from \citet{Jeffries14}. Bottom right: Colour-magnitude diagram of the stars that fall in the proper motion box defined in top right panel.} \end{center}
\end{figure*}

Without any kind of selection beyond the applied quality filters and parallax cuts, the proper motion diagram of the region (Fig.~\ref{fig:fourpanels_regionmap}, top right panel) already shows a significant level of substructure, with two main groups corresponding to the two kinematical populations identified by \citet{Damiani17}. We exclude a priori  the known nearby OCs that are older than the stars of the Gamma Velorum cluster (see HR diagrams in Fig.~\ref{fig:6cmds}) by only keeping the stars with $-8 < \mu_{\alpha*} < -3$\,mas\,yr$^{-1}$  and $5 < \mu_{\delta} < 12$\,mas\,yr$^{-1}$. We include BH~23 in this selection, whose age and proper motion suggest it could be related to the Vela~OB2 stars \citep[also mentioned by][]{Conrad17}, although it is located 50\,pc further away than Gamma Velorum.

\subsection{Photometric selection} \label{sec:photometricselection}

\begin{figure}[ht]
\begin{center} \resizebox{\hsize}{!}{\includegraphics[scale=0.5]{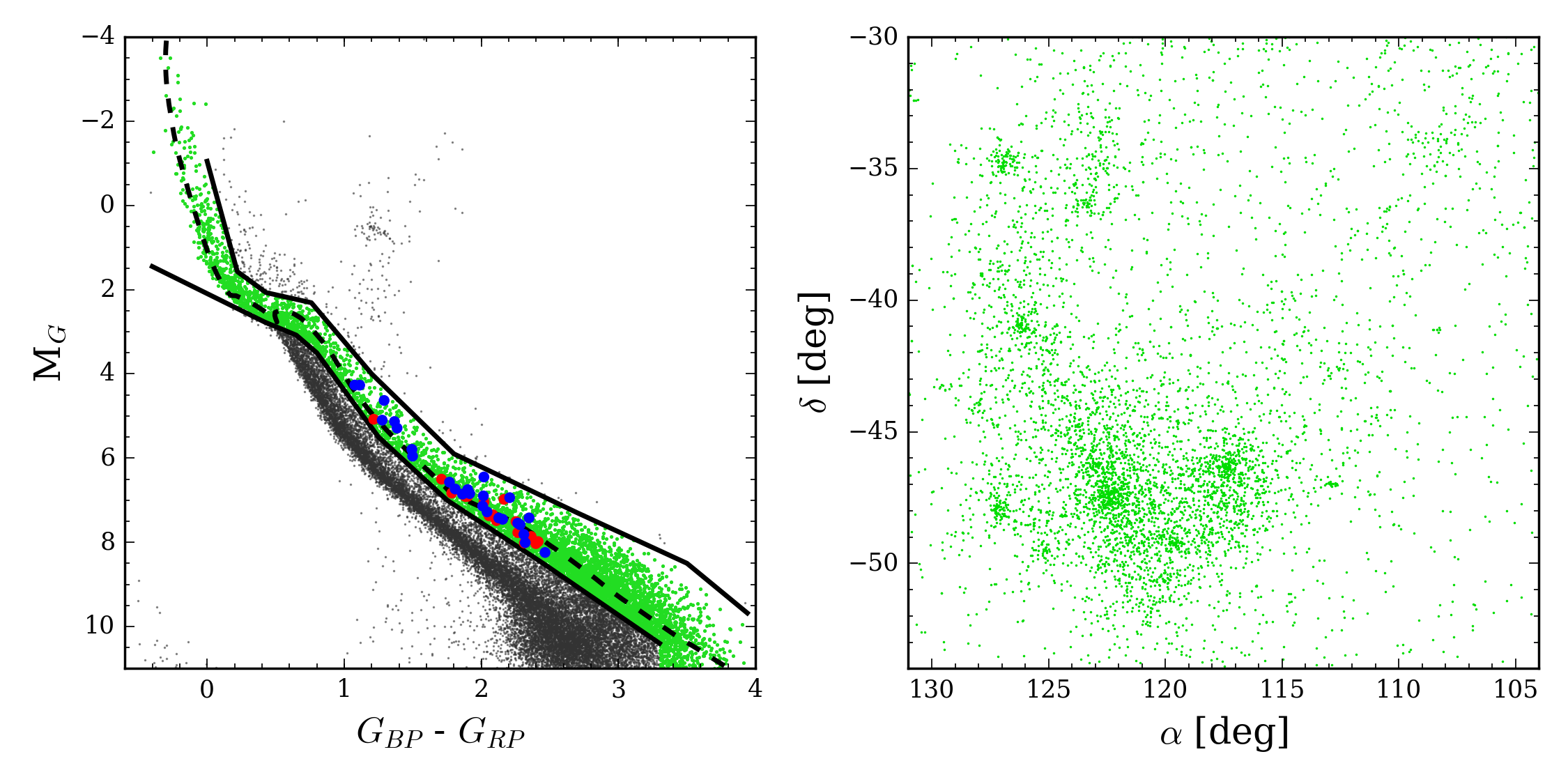}} \caption{\label{fig:photoselection} Left: HR diagram of the stars selected from their proper motions (see Fig.~\ref{fig:fourpanels_regionmap}), showing our photometric selection. The dashed line is a PARSEC isochrone of age 10\,Myr (solar metallicity, $A_V$=0.35). Red and blue points respectively indicate populations A and B from \citet{Jeffries14}. Right: Spatial distribution of the selected stars. } \end{center}
\end{figure}

We then performed a photometric selection meant to separate the clearly visible PMS seen in the bottom right panel of Fig.~\ref{fig:fourpanels_regionmap}. The shape of the applied cuts and the resulting spatial distribution of these stars are shown in Fig.~\ref{fig:photoselection}. The dashed line in Fig.~\ref{fig:photoselection} is a PARSEC isochrone \citep{Bressan12} using the \textit{Gaia} passbands derived by \citet{Evans18}. The choice of isochrone parameters (10\,Myr, solar metallicity, $A_V$=0.35) is not the result of a best fit, but simply highlights the expected path of the PMS in the HR diagram. The lower edge of the filter roughly follows a 30\,Myr isochrone, and is intended to remove the PMS stars of the age of NGC~2547 or older. We verify in Fig.~\ref{fig:HRDseparatecomponents} that our photometric selection produces the desired effect. We note that after performing the photometric selection, the remaining stellar distribution does not extend beyond $\alpha\sim130^{\circ}$ and further south than $\delta\sim-55^{\circ}$, but extends almost 15$^{\circ}$ north of $\gamma^2$~Vel.

\subsection{Cleaning the sample}

We applied the unsupervised classification code UPMASK \citep{KroneMartins14}. The approach does not assume a structure or distribution in spatial or astrometric space, but requires that sources with similar astrometry are more tightly distributed in ($\alpha$,$\delta$) than a uniformly random distribution. The tightness of the distribution is estimated through the total length of the minimum spanning tree (MST) connecting the selected sources, which we require to be shorter than the mean MST of random distributions by at least 3-$\sigma$. The procedure is repeated 100 times, each time shuffling the data within the nominal uncertainties. For each source, the frequency $p$ (ranging from 0\% to 100\%) with which it is classified as a part of a clustered group can be interpreted as a membership probability, or a quantification of the coherence between astrometric ($\varpi$, $\mu_{\alpha*}$, $\mu_{\delta}$) and spatial distribution ($\alpha$, $\delta$). We refer to \citet{CantatGaudin18tgas} for an application of UPMASK to astrometric data. 

\begin{figure}[ht]
\begin{center} \resizebox{\hsize}{!}{\includegraphics[scale=0.5]{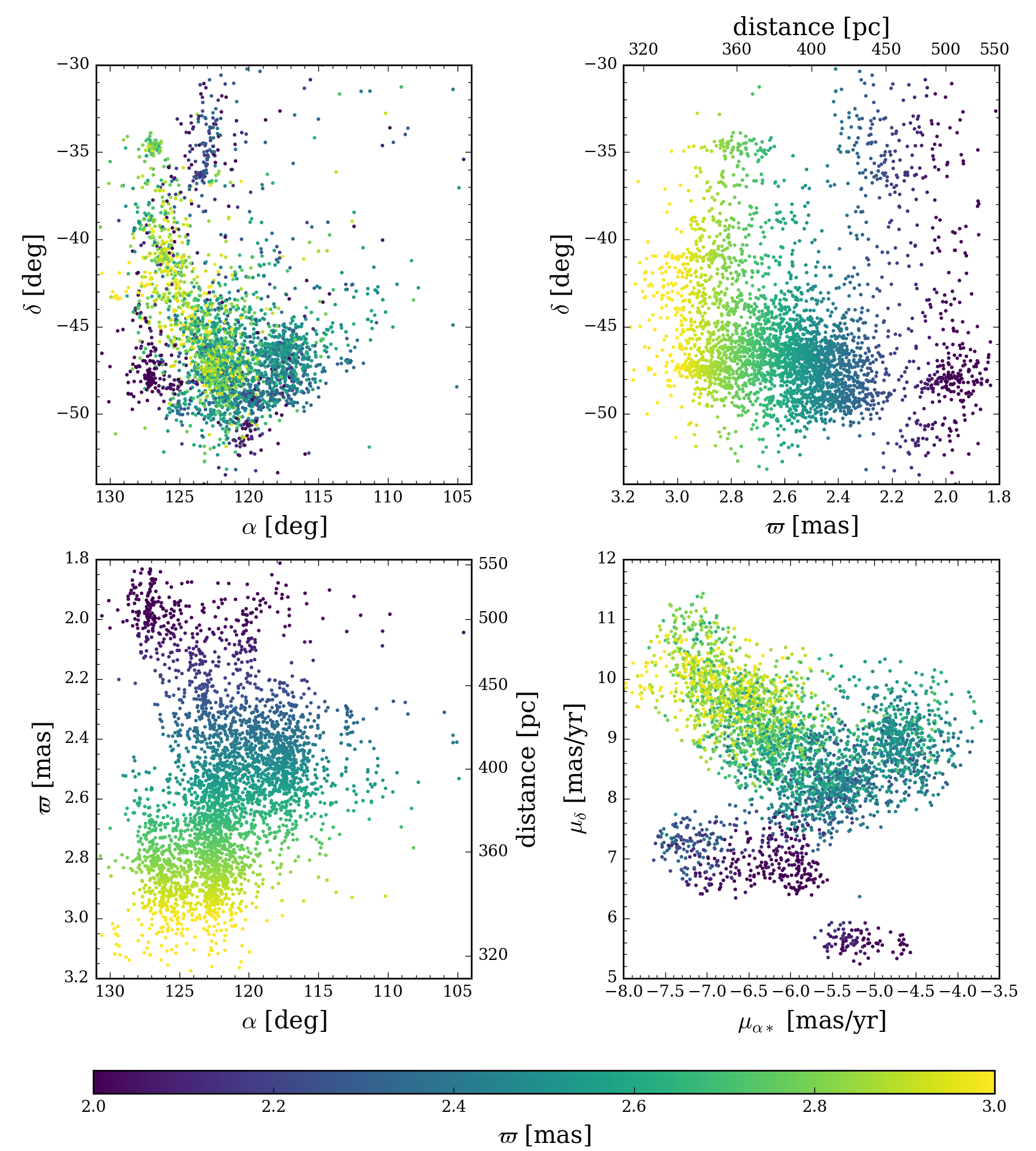}} \caption{\label{fig:mapcolouredbyparallax} All panels use the same colour-coding by parallax. Top left: Distribution of the stars retained as probable members ($p>50\%$) of the Vela~OB2 complex. Top right: $\varpi$ vs $\delta$ for the same stars. Bottom left: $\alpha$ vs $\varpi$. Bottom right: Proper motions. } \end{center}
\end{figure}

The advantage of this approach is that we do not require clusters to be locally dense, but simply that they occupy an  area smaller than the total field of view, allowing  extended or elongated structures to be retained if they are made up of stars with consistent proper motions and parallaxes. The approach is also independent of the number of clusters present in the data, as it only discards the sources that do not appear to belong to any cluster. 
The distribution of sources with $p>50\%$ is shown in Fig.~\ref{fig:mapcolouredbyparallax}. 

Although UPMASK does not rely on assumptions on the spatial distribution of sources (e.g. by fixing a scale or requiring that clusters are centrally concentrated), the procedure still requires the user to arbitrarily define a threshold for the statistical significance of the identified groups, here set to 3-$\sigma$. We note that the loose group seen around \mbox{($\alpha$,$\delta$)=(109$^{\circ}$,-34$^{\circ}$)} in Fig.~\ref{fig:photoselection} shows a large dispersion in proper motions and parallax, and therefore is not classified as significantly clustered by our procedure. These stars could be the remnant of a rapidly expanding group.

The list of stars selected from photometry and the values $p$ we computed are available in the electronic version of this paper.

\section{A fragmented spatial and kinematic distribution} \label{sec:components}

\begin{figure*}[ht]
\begin{center} \resizebox{\hsize}{!}{\includegraphics[scale=0.5]{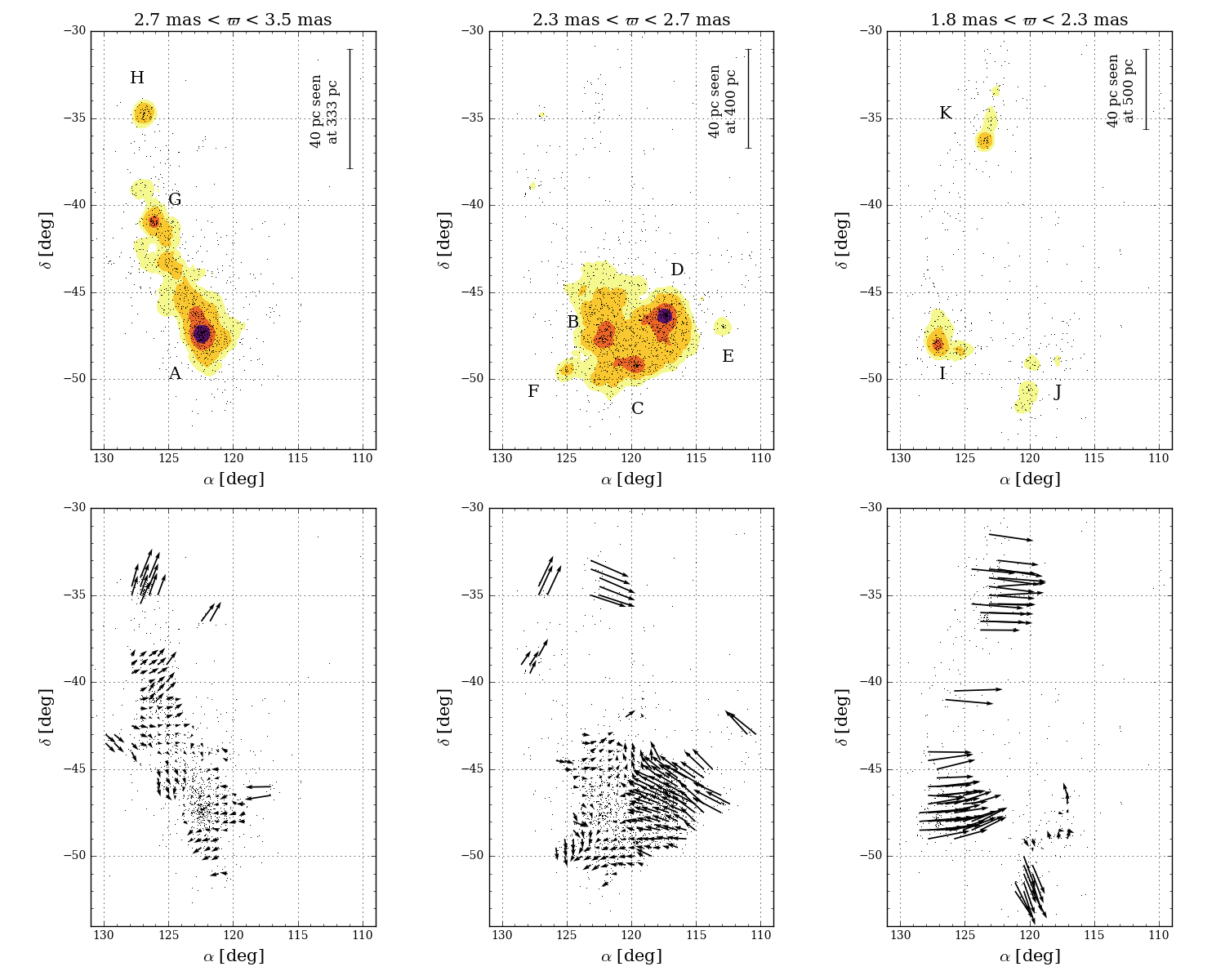}} \caption{\label{fig:contourmap} Top: Spatial distribution of stars in three different parallax ranges, with contour levels showing local surface density in a 0.5$^{\circ}$ radius. Bottom: Length and orientation of arrows indicate the mean tangential velocity (with respect to the main component A) in a 0.5$^{\circ}$ radius, if at least five stars are present within this radius.} \end{center}
\end{figure*}

The stars coeval with the Gamma Velorum cluster are distributed over a large area, with some clearly visible compact clumps (shown in top row of Fig.~\ref{fig:contourmap}). Each of these components exhibits a distinct parallax and proper motion. We arbitrarily label in Fig.~\ref{fig:contourmap} those that appear more prominent. Most of them can be easily distinguished from their location on the sky, but the groups we labelled A, B, C, and D appear continuous and partially overlap in positional space and  in astrometric space. We disentangled their position in astrometric space by modelling their ($\mu_{\alpha*}$,$\mu_{\delta}$,$\varpi$) distribution with a mixture of four covariant Gaussians. 

\begin{figure*}[ht]
\begin{center} \includegraphics[scale=0.55]{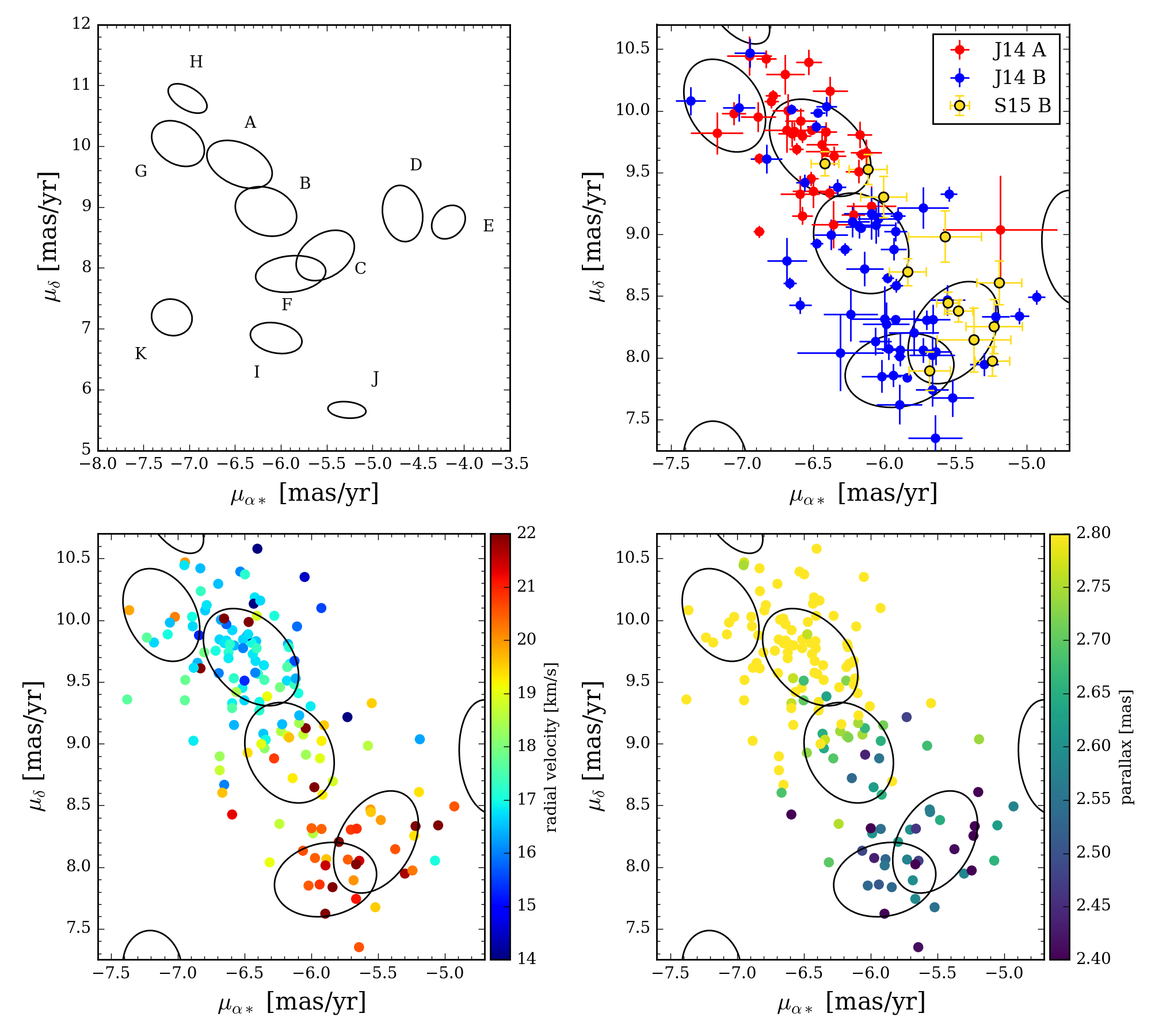} \caption{\label{fig:pm_gmm} Top left: Ellipses represent the mean and standard deviation of proper motions of the main components (as defined in Fig.~\ref{fig:contourmap}). Top right: Proper motions of the stars belonging to population A and B of J14 \citet{Jeffries14} (red and blue, respectively) and population B of \citet{Sacco15} (yellow). Bottom left: Proper motions of all stars with radial velocities in J14 and S15, colour-coded by radial velocity. Bottom right: Same as previous panel, colour-coded by \textit{Gaia}~DR2 parallax.} \end{center}
\end{figure*}

One could arbitrarily define more groupings as individual components, for instance along the bridge of stars connecting our components A and G. The locations and mean astrometric parameters (and associated standard deviations) of the 11 components we define are listed in Table~\ref{table:components}. Their proper motions are shown in Fig.~\ref{fig:pm_gmm}. The proper motion distribution appears continuous for components H, G, A, B, C, and F, but some of the other groups we define exhibit distinct proper motions, in particular the three most distant groups I, J, and K. The HR diagrams of each component are shown in Fig.~\ref{fig:HRDseparatecomponents}.

\begin{figure}[ht]
\begin{center} \resizebox{\hsize}{!}{\includegraphics[scale=0.5]{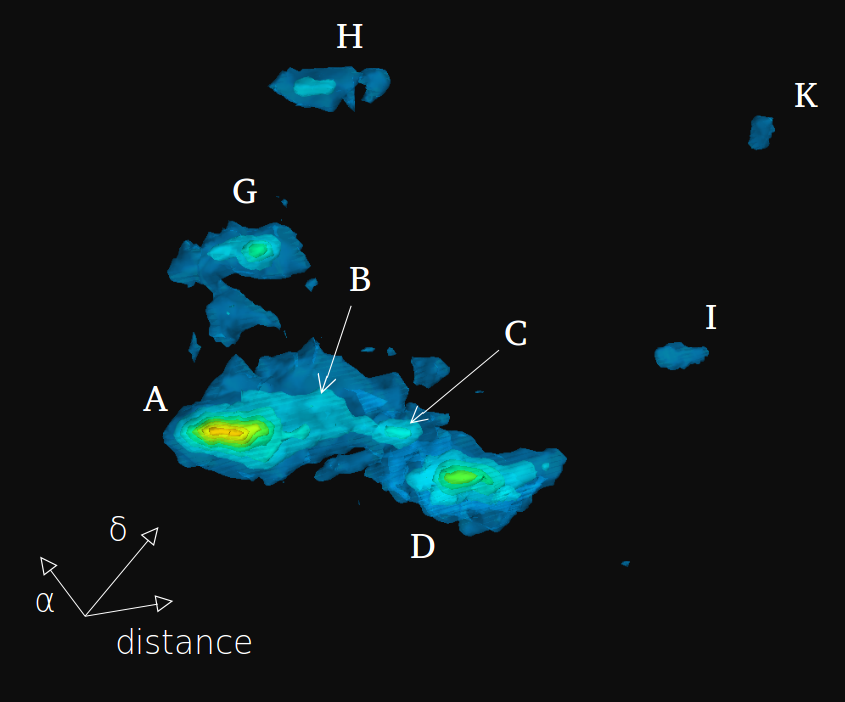}} \caption{\label{fig:view_from_side_2} Isosurfaces of the 3D density distributions of the stars shown in Fig.~\ref{fig:mapcolouredbyparallax}. The distribution was smoothed with a Gaussian kernel of radius 2\,pc. The lower density level shown is $0.5$\,stars\,pc$^{-3}$. Distance from us roughly increases from left to right. The $\alpha$, $\delta$, and distance arrows correspond to physical dimensions of 20\,pc. Components E and J are too sparse to be visible with the adopted contour levels, and component F is behind A and B from this vantage point. The animated figure is available in the electronic version of this article.} \end{center}
\end{figure}

In our final sample, half the stars have fractional parallax errors under 3\%, and 96\% have fractional errors under 10\%. Such small fractional errors allow us to directly estimate distances with the approximation that $d\simeq\frac{1}{\varpi}$. The \textit{Gaia}~DR2 parallaxes are affected by a small negative zero-point offset of about 0.03\,mas \citep{Arenou18}, which for our sources is smaller than the typical nominal uncertainty. Local biases also exist, possibly reaching 0.1\,mas \citep{Lindegren18}, which corresponds to a $\sim$30\,pc bias for the most distant stars in our sample, and to $\sim$10\,pc for the most nearby. Figure~\ref{fig:view_from_side_2} offers a representation of the three-dimensional stellar density distribution, obtained converting the individual parallaxes of each star to distances.

We converted the proper motion of each source to a tangential velocity\footnote{$v_t \simeq 4.74 \frac{\mu}{\varpi}$ with $\mu$ in mas\,yr$^{-1}$, $\varpi$ in mas, $v_t$ in km\,s$^{-1}$.} and plot the velocity relative to the mean of component A in the bottom row of Fig.~\ref{fig:contourmap}. We note that the velocity field is far from homogeneous.

\begin{table*}
\begin{center}
        \caption{ \label{table:components} Mean parameters of the main components identified in Fig.~\ref{fig:contourmap}.}
        \small\addtolength{\tabcolsep}{-1pt}
        \begin{tabular}{c c c c c c c c c c c}
        \hline
        \hline
   component    & $\alpha$      & $\delta$      & $\varpi$ & $\sigma_{\varpi}$ & $\mu_{\alpha*}$ & $\sigma_{\mu_{\alpha*}}$    & $\mu_{\delta}$  & $\sigma_{\mu_{\delta}}$   & RV & $\sigma_{\mathrm{RV}}$\\
                & [deg]         & [deg]         & [mas]    & [mas]   & [mas\,yr$^{-1}$] & [mas\,yr$^{-1}$] & [mas\,yr$^{-1}$] & [mas\,yr$^{-1}$] & [km\,s$^{-1}$] & [km\,s$^{-1}$] \\
        \hline

 A & 122.4 & -47.4   & 2.86 & 0.11 & -6.45 & 0.38 & 9.70 & 0.35 &   16.84 & 0.56   \\
 B & 122.0 & -47.4   & 2.65 & 0.13 & -6.16 & 0.38 & 8.93 & 0.36 &   18.93 & 0.84   \\
 C & 119.7 & -49.4   & 2.45 & 0.11 & -5.52 & 0.35 & 8.20 & 0.36 &   20.85 & 0.90   \\
 D & 117.4 & -46.4   & 2.50 & 0.09 & -4.67 & 0.25 & 8.89 & 0.40 &  - & -    \\
 E & 112.9 & -47.0   & 2.39 & 0.07 & -4.17 & 0.18 & 8.75 & 0.28 &  - & -    \\
 F & 125.0 & -49.5   & 2.47 & 0.12 & -5.89 & 0.38 & 7.90 & 0.30 &  - & -  \\
 G & 126.3 & -40.9   & 2.89 & 0.11 & -7.12 & 0.29 & 10.04 & 0.38 &  - & -    \\
 H & 126.9 & -34.7   & 2.79 & 0.08 & -7.02 & 0.21 & 10.78 & 0.24 &  - & -    \\
 I & 127.1 & -48.0   & 1.97 & 0.07 & -6.05 & 0.28 & 6.85 & 0.25 &  - & -    \\
 J & 120.1 & -50.5   & 2.07 & 0.06 & -5.28 & 0.21 & 5.67 & 0.14 &  - & -    \\
 K & 123.4 & -36.3   & 2.24 & 0.10 & -7.19 & 0.22 & 7.19 & 0.30 &  - & - \\
        \hline
        \hline
        \end{tabular}
\tablefoot{The radial velocity RV is the median of the combined radial velocities from J14 and S15. The associated standard deviation $\sigma_{\mathrm{RV}}$ is 1.4826~$\times$~the median absolute deviation (MAD). }
\end{center}
\end{table*}

Following the choice of J14, we label the Gamma Velorum cluster (surrounding $\gamma^2$~Vel) component A. The compact grouping of a dozen stars located $~1^{\circ}$ north of $\gamma^2$~Vel and labelled cluster 5 by \citet{Beccari18} shares similar proper motions with A, and does not appear as a distinct clump in the density map of Fig.~\ref{fig:contourmap} due to our choice of smoothing length. The mean parallax $\varpi=2.86$\,mas is close to the value of $\varpi=2.895$\,mas quoted by \citet{Franciosini18}, who focused on the stars identified as members of A by J14.

The clump we label B corresponds to the group B of J14, although we note  that some of the stars observed by J14 belong to a third population we call C \citep[cluster 2 of][]{Beccari18}, located $~3^{\circ}$ to the south-west  of $\gamma^2$~Vel. In Fig.~\ref{fig:pm_gmm} (second, third, and fourth panels) we show that the J14 B stars exhibit bimodality in proper motion, parallax, and radial velocity. The J14 B stars with the larger radial velocities also tend to occupy the south of the field investigated by J14. We therefore consider that B and C are distinct objects with RV dispersions of 0.84 and 0.90\,km\,s$^{-1}$ (respectively), rather than one single group with a dispersion of 1.60\,km\,s$^{-1}$ (as reported by J14).
In Fig.~\ref{fig:pm_gmm} we can see that the stars identified by S15 as probable members of Gamma Velorum B belong to this population C. The group we label F can also be found in the southern part of the region, and has similar proper motions to C.

The most massive clump after A is the one we label D. Its location matches the group labelled Escorial~25 by \citet{Caballero08} in their study of OB association with Hipparcos data. This group also corresponds to cluster 1 of \citet{Beccari18}, and is visible in the PMS distribution map by \citet{Armstrong18}. Its tangential velocity in the $\alpha$ direction is larger than the velocity of A, meaning it is reducing its angular distance to A (see bottom middle panel of Fig.~\ref{fig:contourmap}). Unfortunately, without an estimate of the radial velocity of this group of stars, we cannot determine whether its path will cross with other components of the association. Further west lies a small compact group we label E, with a similar tangential velocity to D.

We note a distribution of stars extending from A to the north, itself fragmented, with two major components we label G and H. The velocity distribution along this branch appears to vary continuously (see bottom left panel of Fig.~\ref{fig:contourmap}), with clump G moving away from A, and clump H at the extreme north of this branch moving away from G. 

In the background, we identify three groups with parallaxes smaller than 2.3\,mas. They also exhibit significantly different velocities from the rest of the association. A loose branch of stars is visible linking components I and K (top right and bottom right panels of Fig.~\ref{fig:contourmap}). The stars in this structure also share a common velocity and are consistently moving westwards (with respect to component A). Component J, on the far southern border of the region under study, exhibits a distinct southward velocity. The HR diagrams shown in Fig.~\ref{fig:HRDseparatecomponents} suggest that components I and K are slightly older than the rest of the Vela~OB2 association (but younger than NGC~2547 and the other young OCs present in the area). We note that component K contains the known OC BH~23 (which we included in our initial proper motion selection, as mentioned in Sect.~\ref{sec:propermotionselection}), and appears to match the group labelled Escorial~27 in \citet{Caballero08}.

\section{Mass distribution in the complex} \label{sec:mass}
\citet{Jeffries14} and \citet{Prisinzano16} have pointed out that the presence of such a massive system as $\gamma^2$~Vel (over 30\,M$_{\odot}$) is puzzling in a cluster that does not seem to contain more than $\sim$100\,M$_{\odot}$. According to the empirical scaling relations of \citet{Weidner10}, clusters hosting such a massive star typically contain a total mass of at least 800\,M$_{\odot}$. The results presented in this study show that the current total mass of the Vela~OB2 complex is much higher than previously thought. 

We performed a simple estimation based on the PARSEC isochrone featured in Fig.~\ref{fig:photoselection} (10\,Myr, $A_V$=0.35, solar metallicity), and converted the observed ($G_{BP}-G_{RP}$) colours to masses. Summing up all the individual masses of the stars we consider astrometric members of the association ($p>50\%$), we obtain a total mass of 2330\,M$_{\odot}$. The total mass content is likely higher if the low-mass stars are included as this calculation only includes the stars we could identify as probable members, which are all brighter than $M_G=10.5$ (corresponding to 0.2\,M$_{\odot}$).

\begin{figure}[ht]
\begin{center} \resizebox{\hsize}{!}{\includegraphics[scale=0.5]{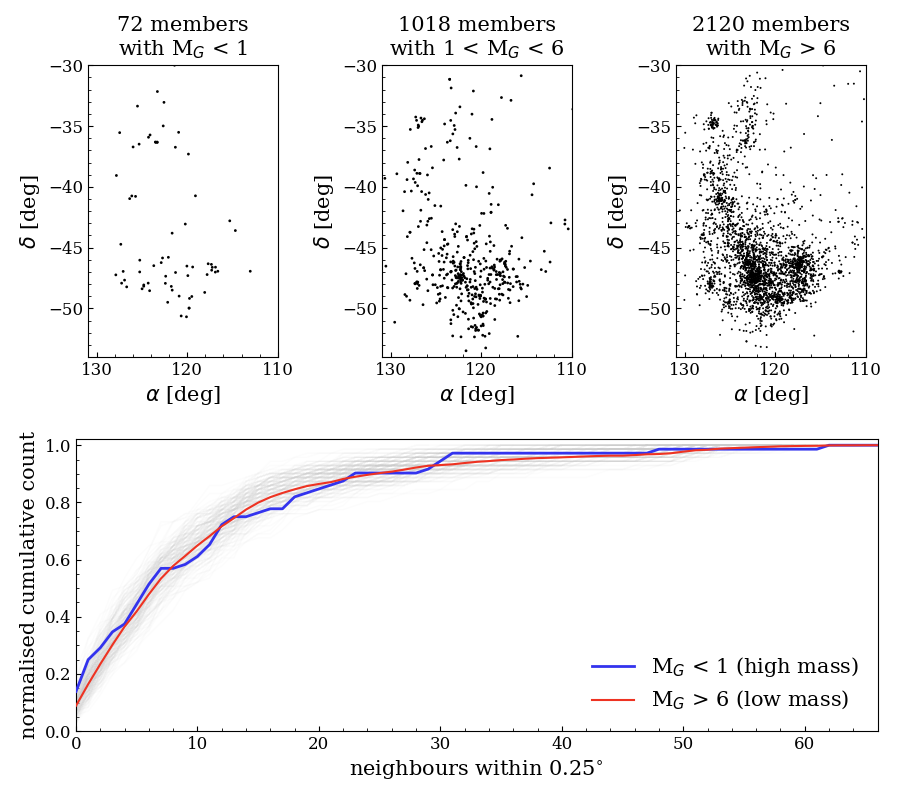}} \caption{\label{fig:mapmassranges} Top row: Spatial distributions of stars in three different absolute magnitude ranges. Bottom panel: Normalised cumulative distribution of the local density (number of neighbours within a fixed radius) for the 72 high-mass stars (blue) and 2120 low-mass stars (red). The grey curves correspond to 200 random re-drawings of 72 points from the total sample.} \end{center}
\end{figure}

We studied the possible difference in spatial distribution between the high-mass and low-mass stars in the complex (shown in the top panel of Fig.~\ref{fig:mapmassranges}). We quantified mass segregation using the approach of \citet{Allison09}, who determine the length of a minimal spanning tree (MST) connecting the brightest stars, and compares it to the MST of a random realisation of the total sample. We also followed the recommendations of \citet{Parker11} and \citet{Maschberger11}, who use moving windows containing fixed numbers of stars. None of these diagnostics was able to reveal a significant level of mass segregation.

We also attempted to quantify whether the massive stars are preferentially distributed in areas that are also occupied by low-mass stars. In this case our results are at odds with those obtained by \citet{Armstrong18}, who observe that the massive stars in their sample do not follow the same spatial distribution as the less massive ones. We performed a similar test to theirs, comparing the local density (considering all stars in the sample) around high-mass and low-mass stars. The cumulative distributions shown in Fig.~\ref{fig:mapmassranges} are essentially identical, with a Kolmogorov–Smirnov test p-value of 0.65. The distribution of the 72 most massive stars is therefore compatible with a random sample of any 72 Vela OB2 stars. We observe that 14\% of the high-mass stars have no neighbour within 0.25$^{\circ}$ (against only 9\% of the low-mass stars).  

Although we cannot explain the discrepancies observed by \citet{Armstrong18} and this study, we note the following:

\begin{itemize}

\item out of the 82 OB stars considered members of Vela~OB2 by \citet{deZeeuw99}, only 61 are present in the \textit{Gaia}~DR2 catalogue, due to the bright magnitude limit of \textit{Gaia};

\item out of these 61 stars, 20 have \textit{Gaia}~DR2 proper motions incompatible with them being members of the Vela~OB2 complex as defined in this study;

\item photometric selections of PMS stars are able to produce purer samples than selections of OB stars, which cannot discriminate between stars of different ages.

\end {itemize}

We therefore conclude that in our sample, the high-mass and low-mass samples do not present statistically distinct distributions.

\section{The large-scale structure and the IRAS Vela Shell} \label{sec:IVS}
\begin{figure*}[ht]
\begin{center} \resizebox{\hsize}{!}{\includegraphics[scale=0.5]{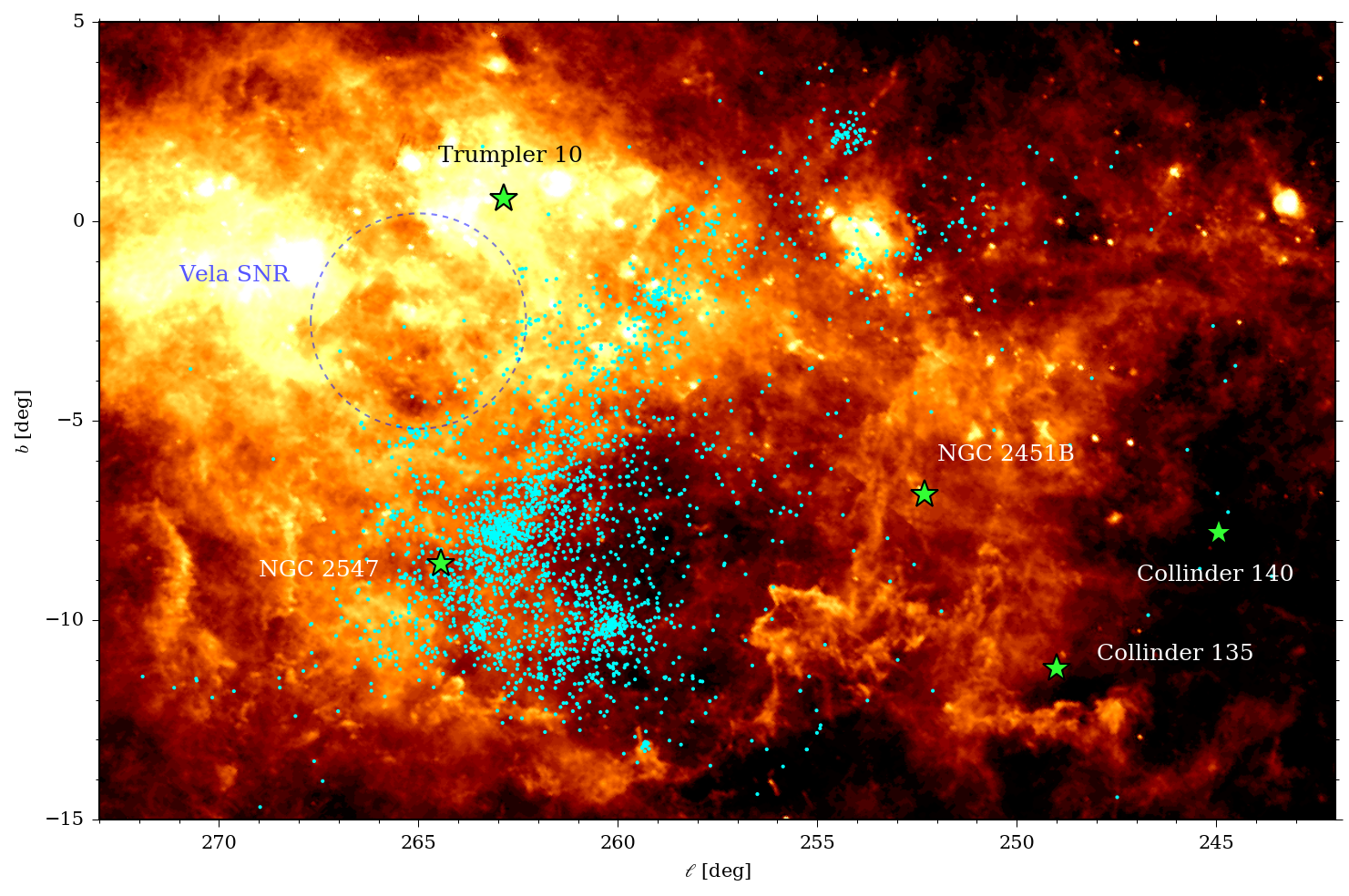}} \caption{\label{fig:IRAS60um} 60\,$\mu$m map from the Improved Reprocessing of the IRAS Survey \citep[IRIS,][]{Miville05}, in Galactic coordinates. The cyan dots are the Vela~OB2 stars identified in this study. The Vela~SNR and the five older OCs located in the region are also indicated.} \end{center}
\end{figure*}

The region studied in this paper is located within the Gum nebula \citep{Gum52}, which has an apparent diameter of more than 30$^{\circ}$ \citep[e.g.][]{Chanot83}. Its origin is still debated, although most authors suggest that the nebula is likely an old supernova remnant \citep{Woermann01,Purcell15}. The Gum nebula is located about 500\,pc from us, straddles the plane of the Milky Way (extending from $l\sim$-15$^{\circ}$ to $l\sim$+15$^{\circ}$), and our line of sight in its direction overlaps with a number of different objects (Fig.~\ref{fig:IRAS60um}). The Vela Molecular Ridge contains a string of HII regions at distances of 800\,pc to 2\,kpc \cite{Rodgers60,May88,Pettersson94,Neckel95,Prisinzano18}, which are not related to the Vela~OB2 complex. They are visible as bright compact regions in Fig.~\ref{fig:IRAS60um}, roughly constrained to $|b|<2^{\circ}$. About $10^{\circ}$ to the north-west of $\gamma^2$~Vel lies the Vela Supernova Remnant (Vela SNR), one of the closest SNRs to us \citep[250$\pm$30,][]{Cha99}, deploying an apparent diameter of $8^{\circ}$. Distance estimates from VLBI parallax place its central pulsar at $287^{+19}_{-17}$\,pc \citep{Dodson03}.

The most relevant structure to our study is the IRAS Vela Shell (IVS). Discovered as a ring-like structure in IRAS \citep{Neugebauer84} far-infrared images by \citet{Sahu92phdt}, and with a proposed radius of $\sim$65\,pc, the IVS appears to be a substructure located near the front face of the Gum nebula. Although not recognised as a distinct structure at the time, the IVS can in fact be seen in the dark cloud maps of \citet{Feitzinger84} \citep[see also][]{Pereyra02}, and traced by the distribution of cometary globules \citep{Zealey79,Zealey83,Reipurth83}. It is also visible in maps of interstellar extinction \citep{Franco12}. \citet{Sridharan92} have suggested that the cometary globule ring bordering the IVS is expanding, a result supported by \citet{Sahu93} and \citet{Rajagopal98}. \citet{Testori06} show that a gas counterpart to the dust IVS exists, and exhibits signs of expansion. \citet{Sushch11} propose a model for the structure of the complex that includes the Vela SNR and the IVS. They suggest that the IVS is the result of the stellar-wind bubble of $\gamma^2$~Vel and the Vela~OB2 association, and that the observed asymmetry of the Vela SNR is due to the envelope of the Vela SNR physically meeting with the IVS. We note that \citet{Woermann01} proposed that the IVS might not be a separate entity, but rather the result of an anisotropic expansion of the Gum nebula.

In this study we find that the spatial distribution of the young stars in the Vela~OB2 complex form a coherent, almost ring-like structure, a hint of which can be perceived in the top right   panel of Fig.~\ref{fig:mapcolouredbyparallax} ($\varpi$ vs $\delta$), and best visualised in Fig.~\ref{fig:view_from_side_2}. The distance from clump A to K is 130\,pc, which matches the early estimates of \citet{Sahu93} for the diameter of the IVS, but the distance from H to I (the most distant clump from us) is 180\,pc, indicating that the ring is not perfectly circular. The relative radial velocities of A, B, and C indicate that their distribution is stretching along our line of sight, as suggested by \citet{Franciosini18}. The relative tangential velocities of A, G, and H indicate that they are being pulled apart as well\footnote{This expansion in the tangential plane is not a virtual effect due to our motion with respect to Vela~OB2, as described in e.g. \citet{Brown97}. If the perspective effect dominated, the positive radial velocity ($\sim$18\,km\,s$^{-1}$) would induce a virtual contraction.}. This behaviour, where all components are moving away from each other, is the signature of an expanding structure. We note that the far side of this ring is not as well-defined as the front side, and with the entire back half only traced by three clumps (I, J, and K) which exhibit slightly different tangential velocities from the rest of the complex (Fig.~\ref{fig:contourmap}), and appear slightly older than the rest of the association as well (see Sect.~\ref{sec:components} and Fig.~\ref{fig:HRDseparatecomponents}).

\section{Discussion} \label{sec:discussion}
The main conclusion of this paper is that the entire stellar distribution of the Vela~OB2 complex is expanding, which indicates that the expansion of the IVS is not due to stellar winds from the young OB stars, but caused by an event that took place before the Vela~OB2 stars were formed, and imprinted the expanding motion into the stellar distribution itself. Their circular distribution hints at an episode of triggered star formation, caused by a mechanism energetic enough to not only compress gas into forming stars, but also induce the expansion of the IVS. \citet{Testori06} have estimated that the total mechanical energy injected into the ISM by the joint action of the massive stars $\gamma^2$~Vel and $\zeta$\,Pup and of the stellar aggregates Trumpler~10 and Vela~OB2 is insufficient by an order of magnitude, and that supernovae might have driven this expansion. The presence of several $\sim$30\,Myr OCs near the IVS (Collinder~135, Collinder~140, NGC~2451B, NGC~2547, Trumpler~10) makes this mechanism possible.

\begin{figure}[ht]
\begin{center} \resizebox{\hsize}{!}{\includegraphics[scale=0.5]{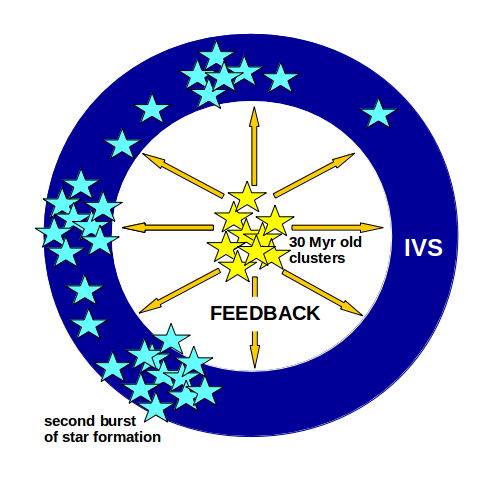}} \caption{\label{fig:cartoon_vela} Schematic picture of the proposed supernova-triggered star formation event. Stellar feedback and supernovae that exploded in the 30\,Myr old clusters (yellow stars) swept the gas, producing the central cavity (white shell) and the IVS (blue shell). The expansion of the IVS triggered a second burst of star formation approximately 10\,Myr ago (cyan stars).} \end{center}
\end{figure}

We propose the following scenario. About 20\,Myr ago, the stars that now form the above-mentioned $\sim$30\,Myr clusters were $\sim$10\,Myr old, corresponding to the lifetime of a $\sim$15\,M$_\odot$ star. Thus, we expect that the region was the site of intense supernova activity, releasing enough energy to sweep gas out of the central cavity and to power the expansion of the IVS (e.g. \citealt{pelupessy2012}). It is reasonable to expect that the expanding IVS produced a shock in the interstellar medium, which triggered a second burst of star formation approximately 10\,Myr ago (e.g. \citealt{machida2005,yasui2006}). According to this scenario, $\gamma{}^2$ Velorum and the other $\sim$10\,Myr old stars were formed in this second burst of star formation. Their location approximately in the rim of the cavity confirms this interpretation. Figure~\ref{fig:cartoon_vela} is a cartoon visualisation of this scenario.

The fact that the OCs in the region are $\sim$30\,Myr old and approximately 20\,Myr older than $\gamma{}^2$ Velorum supports our scenario, because the peak of core-collapse supernova activity occurs in the first $10-20$\,Myr and then it declines fast \citep[see e.g. Figure~1 of][]{Mapelli2013}. This age spread is also compatible with what is observed in numerical simulations of supernova-driven star formation \citep{Padoan16}. No core-collapse supernovae are expected $>50$\,Myr after cluster formation.

\citet{Testori06} mention that \citet{Hoogerwerf01} identified a B-type runaway star (HD~64760, HIP~38518), and suggested it was ejected in a binary-supernova scenario that took place in the Vela OB2 region. The identification of hypervelocity stars from \textit{Gaia} data \citep{Renzo18,Marchetti18,MaizApellaniz18,Irrgang18} might tell us whether other such objects are compatible with an origin inside the IVS. A full dynamical modelling of the orbits of hypervelocity stars and nearby open clusters is beyond of the scope of this study, but should be possible based on the 3D velocities of these objects \citep{Soubiran18}. The fact that Trumpler~10, NGC~2451B, Collinder~135, Collinder~140, and NGC~2547 all have very similar ages (see Fig.~\ref{fig:6cmds}) even suggests that they too might have originated from a single event of triggered star formation. The sparse group reported by \citet{Beccari18} in the foreground of the region also shares a common age with these five  clusters.

The supernova scenario also  accounts for the slight age difference between groups I, J, and K (as labelled in this study) and the rest of the Vela~OB2 complex. If the gas distribution was not perfectly homogeneous at the time of the explosion, the compression wave caused by the supernova might have hit denser areas at the back of the bubble and initiate star formation there a few Myr before star formation took place in the front face of the shell. For a current diameter from 130\,pc \citep[the distance from A to K, in agreement with the early estimate of][]{Sahu93} to 180\,pc (the distance fro H to I), assuming the supernova event took place 10--20\,Myr ago yields a constant expansion speed of the shell of 13 to 18\,km\,s$^{-1}$ (6.5 and 9\,km\,s$^{-1}$ with respect to the centre, respectively), in agreement with the 13\,km\,s$^{-1}$ value of \citet{Rajagopal98} and with the 15.4\,km\,s$^{-1}$ of \citet{Testori06}.

In this study we did not attempt to characterise the dynamical state of each of the observed components. However, we have shown that the total mass of the structure is much larger than previously thought. We have also shown that the internal velocity dispersion of component B is smaller than the value of 1.6\,km\,s$^{-1}$ assumed by J14 and \citet{Mapelli15}. This smaller velocity dispersion in a deeper potential well indicates that B is not as supervirial as previously thought. A dedicated study of each component could determine whether any of them is going to survive as a bound cluster, or whether all of them will eventually disrupt. The typical expansion speed observed by \citet{Kuhn18} in clusters younger than 5\,Myr is $\sim$0.5\,km\,s$^{-1}$, corresponding to $\sim$0.5\,pc\,Myr$^{-1}$, which for a constant expansion rate leads to physical sizes of 5\,pc after 10\,Myr, matching the typical size of the observed components of Vela~OB2. 

The exact age of the Vela~OB2 stars is still a debated issue;  determinations from CMDs suggest an age between $\sim$7.5\,Myr \citep{Jeffries17} and $\sim$10\,Myr \citep{Prisinzano16}, while lithium depletion measured from spectroscopy are compatible with an age of $\sim$20\,Myr \citep{Jeffries17}. The discrepancy between the ages of PMS stars obtained from lithium depletion patterns and from photometry has been noted in the literature \citep[e.g.][]{Bell13,Bell14,Messina16,Bouvier18}. Our scenario relies on the age difference between the various populations rather than on their absolute age, and still holds if the photometric ages of the younger and older populations are underestimated by a similar amount. The proposed scenario would only be invalid if the age difference turned out to be larger than $\sim50$\,Myr, as the last core-collapse supernovae from the older population would have happened long before the formation of the younger population.

\section{Conclusion} \label{sec:conclusion}
In this paper we make use of the \textit{Gaia}~DR2 astrometry and photometry to identify the young stars of the Vela~OB2 association over more than 100\,pc. We identify and list 11 main components of a fragmented distribution. They appear to roughly follow an expanding ring-like structure. 

We observe a good morphological correlation between their three-dimensional distribution and kinematics and the dust and gas structure known as the IRAS Vela Shell. The expanding stellar distribution indicates that the expansion of the shell is not primarily due to the Vela~OB2 stars, but to an event that impacted the local dynamics before the Vela~OB2 stars formed. This observation is compatible with previous suggestions that the shell originates from a supernova explosion. The progenitor of this supernova might have belonged to one of the several clusters with ages $\sim$20\,Myr older than Vela~OB2 located in the region.

The lack of radial velocities for all stars in the association prevents us from painting a full three-dimensional dynamic portrait of the region and its expansion. Robust age determinations and orbits for all the populations in the region are necessary in order to establish a physically plausible sequence of events and prove or disprove the hypothesis that the formation of the Vela~OB2 stars was triggered by the supernova that initiated the expansion of the IRAS Vela Shell.

\section*{Acknowledgements}

TCG acknowledges support from the Juan de la Cierva - formaci\'on 2015 grant, MINECO (FEDER/UE). This work was supported by the MINECO (Spanish Ministry of Economy) through grant ESP2016-80079-C2-1-R (MINECO/FEDER, UE) and MDM-2014-0369 of ICCUB (Unidad de Excelencia `María de Maeztu'). MM acknowledges financial support from the MERAC Foundation, from INAF through PRIN-SKA, from MIUR through Progetto Premiale `FIGARO' and `MITiC', and from the Austrian National Science Foundation through the FWF stand-alone grant P31154-N27. 

The preparation of this work has made extensive use of Topcat \citep{Taylor05}, and of NASA's Astrophysics Data System Bibliographic Services, as well as the open-source Python packages Astropy \citep{Astropy13}, numpy \citep{VanDerWalt11}, scikit-learn \citep{scikit-learn}, and Mayavi \citep{ramachandran2011mayavi}.
The figures in this paper were produced with Matplotlib \citep{Hunter07}. This research has made use of Aladin sky atlas developed at CDS, Strasbourg Observatory, France \citep{Bonnarel00aladin,Boch14aladin}. 

This work presents results from the European Space Agency (ESA) space mission Gaia. Gaia data are being processed by the Gaia Data Processing and Analysis Consortium (DPAC). Funding for the DPAC is provided by national institutions, in particular the institutions participating in the Gaia MultiLateral Agreement (MLA). The Gaia mission website is \url{https://www.cosmos.esa.int/gaia}. The Gaia archive website is \url{https://archives.esac.esa.int/gaia}.


\bibliographystyle{aa} 
\linespread{1.5}                
\bibliography{biblio}

\clearpage
\appendix

\section{HR diagrams of nearby open clusters}

\begin{figure*}[!ht]
\begin{center} \includegraphics[scale=0.45]{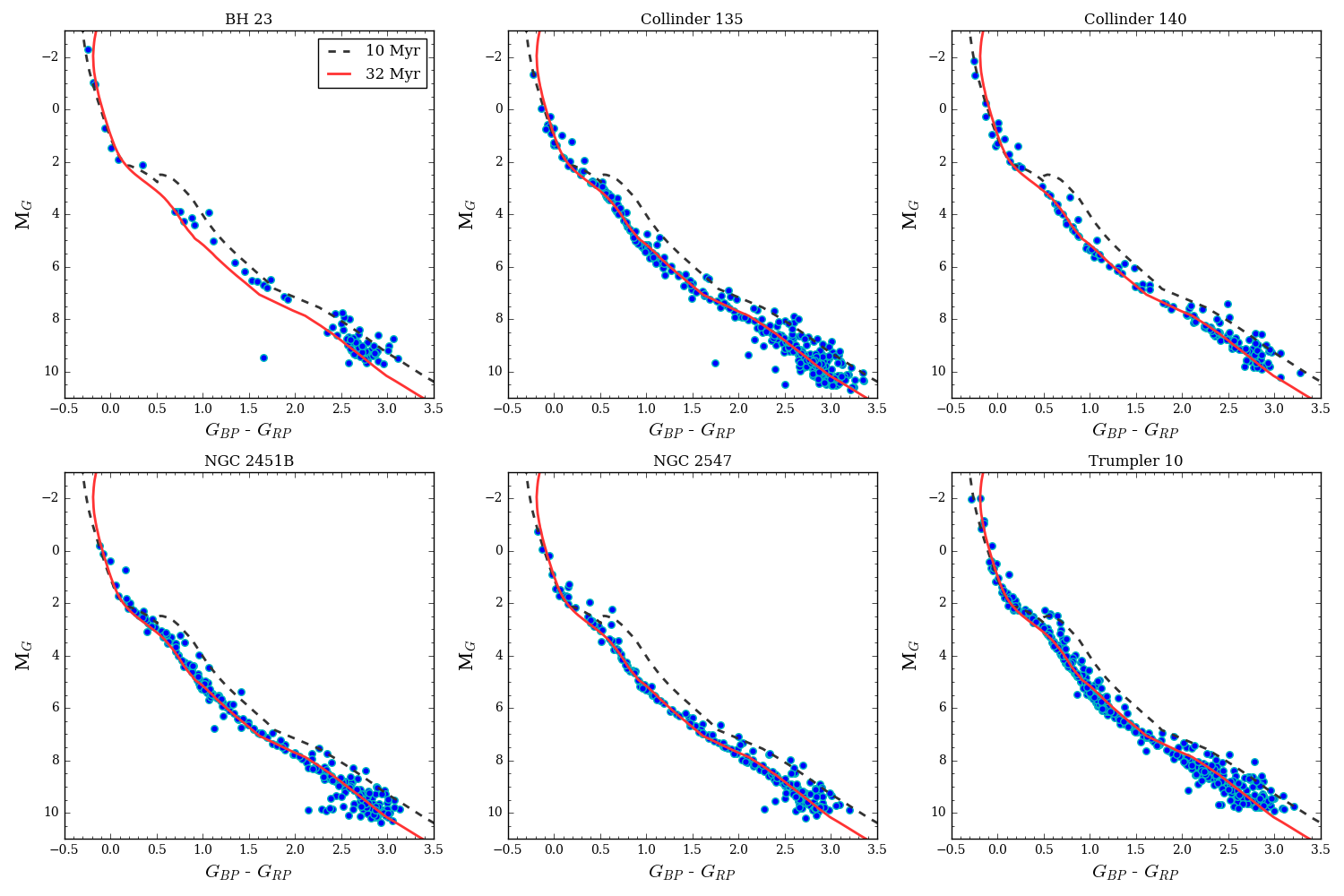} \caption{\label{fig:6cmds} HR diagrams of known OCs in the area of the Vela~OB2 association. The membership was taken from \citet{CantatGaudin18gdr2} and is based on astrometry only. The dashed and continuous lines correspond to PARSEC isochrones of 10 and 32\,Myr, respectively; $A_V$=0.35; and solar metallicity.} \end{center}
\end{figure*}

\clearpage
\section{HR diagrams of individual components}

\begin{figure*}[!ht]
\begin{center} \includegraphics[scale=0.4]{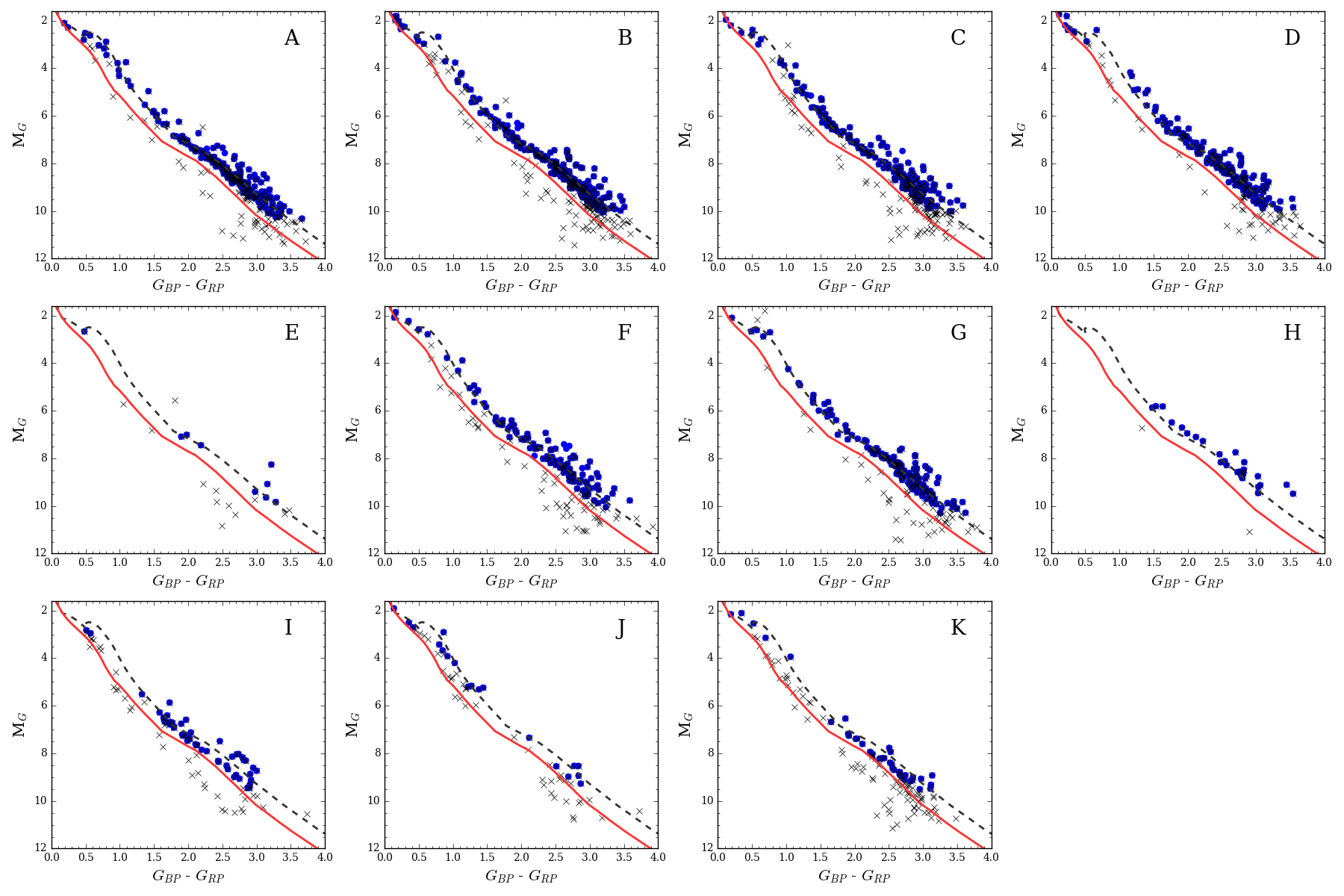} \caption{\label{fig:HRDseparatecomponents} HR diagrams of the main identified components. Stars only compatible on the basis of their astrometry (black crosses); stars that also pass the photometric selection defined in Sect.~\ref{sec:photometricselection} (blue dots). The dashed and continuous lines correspond to PARSEC isochrones of 10 and 32\,Myr, respectively;  $A_V$=0.35; and solar metallicity.} \end{center}
\end{figure*}


\end{document}